\documentclass[prb,twocolumn,showpacs,nonpreprintnumbers,amsmath,amssymb]{revtex4}

\usepackage{graphicx}
\usepackage{dcolumn}
\usepackage{bm}
\usepackage{subfigure}

\begin{document}

\title{Ground State Properties of one-dimensional Antiferromagnetic Spin-1 Chain with Single-ion Anisotropy}
\author{Ekrem Ayd\i ner}
\email{ekrem.aydiner@deu.edu.tr}
 \author{Cenk Aky\"{u}z}\affiliation{Department of Physics, Dokuz Eyl\"{u}l University \\Faculty of Arts and Sciences
Tr-35160 Izmir, Turkey}
\date{\today}

\begin{abstract}
In this study, we have investigated ground state properties of
one-dimensional antiferromagnetic spin-1 chain with single-ion
anisotropy at very low temperatures using the Transfer Matrix
method. Magnetic plateaus, phase diagram, specific heat,
susceptibility of the spin chain have been evaluated numerically
from the free energy. Results are in good agreement with the
experimental data for the spin-1 compounds
[Ni$_2$(Medpt)$_2$($\mu$-ox)(H$_2$O)$_2$](ClO$_4$)$_2$2H$_2$O,
[Ni$_{2}$(Medpt)$_{2}$($\mu$-ox)($\mu$-N$_{3}$)](ClO$_{4}$)0.5H$_{2}$O,
Ni(C$_2$H$_8$N$_2$)Ni(CN)$_4$ and
Ni(C$_{10}$H$_8$N$_2$)$_2$Ni(CN)$_4$.H$_2$O. However, spin-Peierls
transition have not been observed in the temperature dependence of
specific heat and magnetic susceptibility.
\end{abstract}


\maketitle The physics of low-dimensional i.e., one-dimensional
(1D) or quasi-one-dimensional (Q1D) spin-S (S$\geq1$) chains has
attracted a considerable amount of attention after the prediction
by Haldane\cite{1} that a 1D Heisenberg antiferromagnet should
have an energy gap between the singlet ground state and the first
excited triplet states in the case of an integer spin quantum
number, while the energy levels are gapless in the case of a half
integer spin quantum number. The most fascinating characteristic
of these low-dimensional systems is that they show magnetic
plateaus i.e., quantization of magnetization at low temperatures
near the ground state. This phenomenon has been observed not only
in Haldane spin systems but also in other spin gapped systems for
instance; spin-Peierls and spin ladders systems. A general
condition of quantization of the magnetization was derived from
the Lieb-Schultz-Mattis theorem \cite{2} for low-dimensional
magnetic systems. The fact that, Oshikawa, Yamanaka and Affleck
(OYA) \cite{3} discussed this plateau problem and derived a
condition $p(S-m)=integer$, necessary for the appearance of the
plateau in the magnetization curve of one-dimensional spin system,
where $S$ is the magnitude of spin, $m$ is the magnetization per
site and $p$ is the spatial period of the ground state,
respectively. The $2S+1$ magnetization plateaus (contained the
saturated magnetization $m=S$) can appear when the magnetic field
increases from zero to its saturation value $h_{s}$. Theoretical
studies have suggested the realization of quantization of
magnetization in various
systems,\cite{3,4,5,6,7,8,9,10,11,12,13,14} and it has been
observed in experimental studies.\cite{15,16,17,18,19}

Recent years, antiferromagnetic (AF) spin-1 systems among the
other low-dimensional gapped spin-S systems have drawn attention
from both theorists and experimentalists in literature. So far,
many Q1D gapped AF spin-1 systems, which are called Haldane
systems, spin-Peierls and ladder compounds, were synthesized and
to understand the ground state behaviors of the spin-1 AF
Heisenberg chain have been extensively studied. The first Q1D
spin-1 Haldane gap compound was
Ni(C$_2$H$_8$N$_2$)$_2$NO$_2$(ClO$_4$) (abbreviated NENP) which
was synthesized by Meyer et al.\cite{20} in 1981, just before the
Haldane's prediction appeared. Renard et al.\cite{21} showed that
NENP had the energy spectrum predicted by Haldane; the magnetic
susceptibility decreased steeply at low temperatures for all of
the crystal axes suggesting a single ground state and neutron
inelastic scattering showed that the magnetic excitation had a gap
at the AF zone center. Other Q1D spin-1 compounds
Ni(C$_2$H$_8$N$_2$)$_2$Ni(CN)$_4$ (abbreviated
NENC),\cite{22,23,24,25} Ni(C$_{11}$H$_{10}$N$_2$O)$_2$Ni(CN)$_4$
(abbreviated NDPK),\cite{22,23}
Ni(C$_{10}$H$_8$N$_2$)$_2$Ni(CN)$_4$.H$_2$O (abbreviated
NBYC),\cite{22,23,24} Ni(C$_5$H$_{14}$N$_2$)$_2$N$_3$(PF$_6$)
(abbreviated NDMAP),\cite{26,27,28} and
Ni(C$_5$H$_{14}$N$_2$)$_2$N$_3$(ClO$_4$) (abbreviated
NDMAZ)\cite{29} have also been identified as effective Heisenberg
chains. This class of compounds exhibits a non-degenerate ground
state which can be separated from the lowest excitation. As
predicted by Haldane, it is experimentally shown that these
systems have an energy gap between the single ground state and
first excited triplet.\cite{21,22,23,24,25,26,27,28,29} In
addition, exception Haldane systems, several Q1D AF spin-1 gapped
compounds such as AF dimer compound
[Ni$_2$(Medpt)$_2$($\mu$-ox)(H$_2$O)$_2$](ClO$_4$)$_2$2H$_2$O,\cite{15}
and AF alternating chain compound
[Ni$_{2}$(Medpt)$_{2}$($\mu$-ox)($\mu$-N$_{3}$)](ClO$_{4}$)0.5H$_{2}$O,
\cite{16} where (Medpt=methyl-bis(3-aminopropyl)amine), the ladder
system 3,3',5,5'-tetrakis (N-tert-butylaminxyl)biphenyl
(BIP-TENO),\cite{17,30,31} the spin-Peierls
systems\cite{32,33,34,35} (Li,Na)V(Si,Ge)$_2$O$_6$ have been
synthesized and magnetic properties have been studied.

The magnetic quantization behavior has been experimentally
observed in only several gapped spin-1 AF materials as predicted
by OYA.\cite{3} For example, Narumi et al.\cite{15,16} observed a
magnetic plateaus in the magnetization curve for both
[Ni$_{2}$(Medpt)$_{2}$($\mu$-ox)(H$_{2}$O)$_{2}$](ClO$_{4}$)$_{2}$2H$_{2}$O
and
[Ni$_{2}$(Medpt)$_{2}$($\mu$-ox)($\mu$-N$_{3}$)](ClO$_{4}$)0.5H$_{2}$O.
Goto et al.\cite{17} reported the existence of the magnetization a
plateau at $0.25$ in spin-1 3,3',5,5'-tetrakis
(N-tert-butylaminxyl)biphenyl (BIP-TENO). On the other hand, in
theoretical studies, Chen et al. \cite{4} employed the classical
Monte Carlo method to investigate the magnetization plateaus of 1D
classic spin-1 AF Ising chain with a single-ion anisotropy under
the external field at low temperatures, and they showed that the
system has $2S+1$ magnetic plateaus. Tonegawa et al. \cite{9}
observed the plateau of at $m=$0.0, 0.5, and 1.0 in the ground
state of spin-1 AF Heisenberg chain with bond alternating and
uniaxial single-ion anisotropy. Sato and Kindo obtained a magnetic
plateau at $m=$0.0, 0.5, and 1.0 for spin-1 bond alternating
Heisenberg chain using renormalization group method.\cite{14}

Other thermodynamical features of low-dimensional AF spin-1
systems have been extensively studied in theoretical concepts as
well as in experimental studies. For example, White and
Huse\cite{36} calculated the ground state energy and Haldane gap
of a spin-1 AF Heisenberg chain using the density matrix
renormalization group techniques. Yamamoto and Miyashita\cite{37}
studied the specific heat and magnetic susceptibility of a spin-1
AF chain by Monte Carlo simulation. The spin-wave theory was used
by Rezende\cite{38} to study the temperature dependence of
excitation spectra, and it has been found that the excitation
energy or Haldane gap increases with increasing temperature in the
region of low temperature. Campana et al.\cite{39} discussed the
specific heat and magnetic susceptibility of a finite chain by a
numerical calculation method. Bao\cite{40} employed a second-order
Green function method to discuss the variations of internal energy
and specific heat with temperature for a spin-1 Heisenberg AF
chain. Li and Zhu,\cite{41} using a self-consistent mean-field
approximation, studied the thermodynamic properties of a spin-1 AF
Heisenberg chain with an anisotropy in the presence of an external
magnetic field at a finite temperature, they found that the
internal energy and Haldane gap increase with increasing
temperature. Batchelor et al.\cite{42} investigated the ground
state and thermodynamic properties of spin-1 via the integrable
$su(3)$ model. Lou et al.\cite{43} calculated the magnetic
susceptibility and specific heat of one-dimensional spin-1
bilinear-biquadratic Heisenberg model using transfer matrix
renormalization group. On the other hand, a field-theoretic
approach to the Heisenberg spin-1 chain with single-ion anisotropy
was suggested by Tsvelik.\cite{44}

Now, it understood from experimental and theoretical studies that
quantization of magnetization appears due to gap mechanism
originating from dimerization, frustration, single-ion anisotropy,
periodic field so on. However, it notes that low-dimensional
gapped spin systems have still open problems and noteworthy
physics. Therefore, in this study, we interested in ground state
properties of 1D AF spin-1 chain.

In general, the most of the theoretical studies about
low-dimensional spin-S systems have been based on Heisenberg
Hamiltonian because of non-trivial quantum effects. However, in
several theoretical studies have been shown that it is possible to
used classical spin systems to obtain magnetic plateaus and other
thermodynamical behaviors.\cite{4,11,13} Considering this fact, we
have constructed herein the Hamiltonian of the 1D spin-1 system in
a quasi-classical manner which allows to use the Ising type
variable instead of the quantum spin operators. Such a Hamiltonian
with single-ion anisotropy for 1D spin-1 systems is described by
\begin{equation}
H=J\sum\limits_{i=1}^{N}S_{i}^{z}S_{i+1}^{z}+D\sum\limits_{i=1}^{N}(S_{i}^{z})^2+h\sum\limits_{i=1}%
^{N}S_{i}^{z} \label{this}
\end{equation}
where $J$ denotes the exchange coupling of antiferromagnetic type
($J>0$), $D$ describe the single-ion anisotropy, and $h$ is the
external field. Also, $S_{i}^{z}$ refers to spin of magnitude $1$,
which takes on $0$, and $\pm1$ values.

We used Transfer Matrix method to obtain ground state properties
of 1D AF spin-1 system which was defined by Eq. (1). Using this
method, it is possible to get the analytical and comparable
results for 1D infinite spin system, though it is very simple
approach.

For the Eq. (1) transfer matrix $V$ is given as
\begin{widetext}
\begin{equation}
V=\left(%
\begin{array}{ccc}
  \exp{(-\beta{J}-\beta{h}-\beta{D})} &  \exp(-\beta{h}/2-\beta{D}/2) & \exp(\beta{J}-\beta{D}) \\
  \exp(-\beta{h}/2-\beta{D}/2) & 1 & \exp(\beta{h}/2-\beta{D}/2) \\
  \exp(\beta{J}-\beta{D}) & \exp(\beta{h}/2-\beta{D}/2) & \exp(-\beta{J}+\beta{h}-\beta{D}) \\
\end{array}%
\right)
\end{equation}
\end{widetext}
where $\beta=1/kT$. Using relation (2), partition function of the
1D spin-1 AF systems can be represented in terms of transfer
matrix $V$ as
\begin{equation}
\mathcal{Z}=Tr \textit{V}^N.
\end{equation}
Hence free energy per spin of this system is given by
\begin{equation}
f(h,T)=-kT\lim\limits_{N\rightarrow\infty}\frac{1}{N}\ln\mathcal{Z}.
\end{equation}
We are interested in the three thermodynamical expressions of (i)
ferromagnetic order $m$, (ii) the specific heat $C$, and (iii) the
magnetic susceptibility $\chi$, which respectively are
\begin{subequations}
\begin{equation}
m=-\frac{\partial{f(h,T)}}{\partial{h}}
\end{equation}
\begin{equation}
C=-T\frac{\partial^2{f(h,t)}}{\partial{T^2}}
\end{equation}
\begin{equation}
\chi=-\frac{\partial^2{f(h,T)}}{\partial{h^2}}.
\end{equation}
\end{subequations}
A conventional way to calculate free energy Eq. (4) which is
required to obtain expression in Eq. (5) is that partition
function is expressed in terms of eigenvalues of transfer matrix
(2) as
\begin{equation}
\mathcal{Z}=Tr\textit{V}^N
=\lambda_{1}^N+\lambda_{2}^N+\lambda_{3}^N.
\end{equation}

The eigenvalues of the Transfer Matrix (2) have been carried out
with the aid of \verb"Mathematica 4.0" package. However, we have
not listed herein since they occupied many pages. After we tested
all eigenvalues numerically, we have sorted and so-called them
from the biggest to smaller as $\lambda_{1}$, $\lambda_{2}$,
$\lambda_{3}$ respectively. Using standard assumptions of Transfer
Matrix method, Eq. (6) was defined as,
\begin{equation}
\mathcal{Z}=Tr\textit{V}^N
=\lambda_{1}^N[1+({\frac{\lambda_{2}}{\lambda_{1}}})^N+
({\frac{\lambda_{3}}{\lambda_{1}}})^N].
\end{equation}
It is clearly seen that on the RHS of Eq.\, (7) the second and the
third terms goes to zero as $N\rightarrow{\infty}$ since both
terms are being ${\mid{\frac{\lambda_{2}}{\lambda_{1}}}\mid}<1$
and ${\mid{\frac{\lambda_{3}}{\lambda_{1}}}\mid}<1$. Consequently,
the free energy Eq.\, (4) is reduced to
\begin{equation}
f(h,T)=-kT\ln{\lambda_1}
\end{equation}
then, some of the thermodynamical quantities of the system, we
interested in Eq. (5), can be calculated numerically using Eq. (8)
which is depend on the biggest eigenvalue $\lambda_1$ of the
matrix (2).
\begin{figure}
\begin{center}
\subfigure[\hspace{0.4cm}]{\label{fig:sub:a}
\includegraphics[width=8.0cm,height=8.0cm,angle=0]{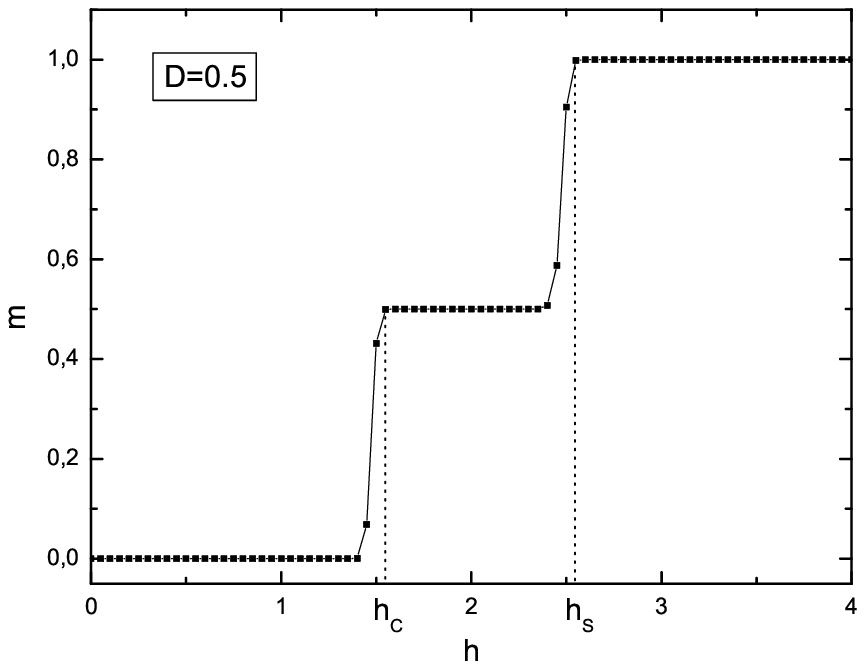}}
\hspace{0.4cm} \subfigure[\hspace{0.4cm}]{\label{fig:sub:b}
\includegraphics[width=8.0cm,height=8.0cm,angle=0]{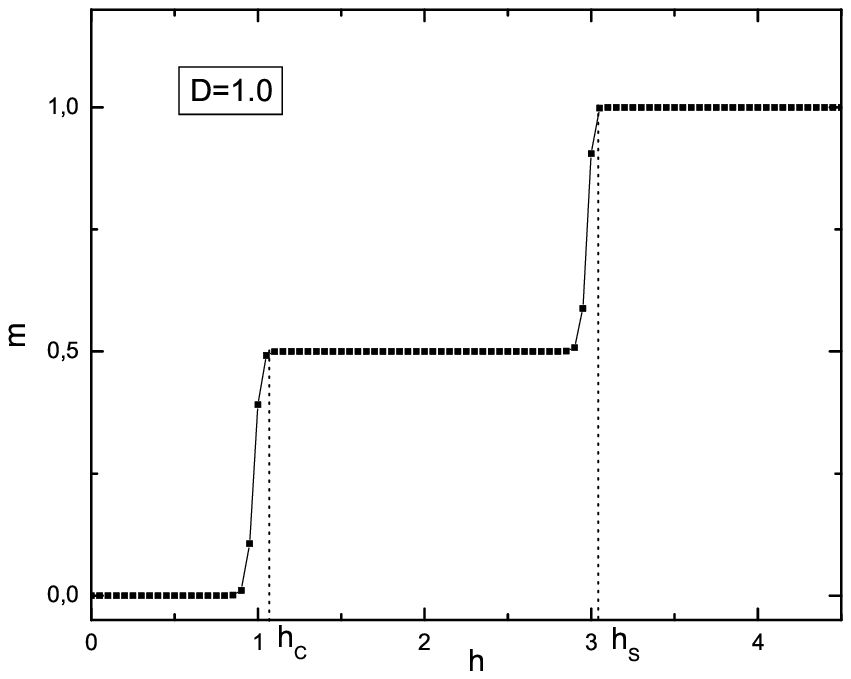}}
\caption{The magnetization $m$ as a function of magnetic field h
for (a) $D=0.5$; and (b) $D=1.0$. Where $h_c$ and $h_s$ are the
initial and the saturated field, respectively. ($T=0.01$, unit by
$J$)} \label{fig:sub:Fig1a-Fig1b}
\end{center}
\end{figure}

Obtained numerical results are followed: In Fig. 1(a) and (b), the
magnetization $m$ is plotted at $T=0.01$ (units by $J$) as a
function of external field $h$ for $D=0.5$, and $D=1.0$,
respectively. It is seen that $2S+1=3$ plateaus appear (i.e.
$m=$0, 0.5, and 1) for all values of positive single-ion
anisotropy $D$ at the low temperature near the ground state under
external field $h$. No magnetization appears in the low field
region. By increasing the field there is a plateau of which
starting point is called initial critical field $h_{c}$ and then
with increasing the field, a plateau of which starting point is
called saturated field $h_{s}$ occurs again. These results clearly
indicated that numbers of plateaus of classical 1D spin-1 AF
system obey to OYA criterion as well as Monte Carlo
prediction,\cite{4} and compatible with the experimental data for
compound
[Ni$_2$(Medpt)$_2$($\mu$-ox)(H$_2$O)$_2$](ClO$_4$)$_2$2H$_2$O.\cite{15}
\begin{figure}
\begin{center}
\includegraphics[width=8.0cm,height=8.0cm,angle=0]{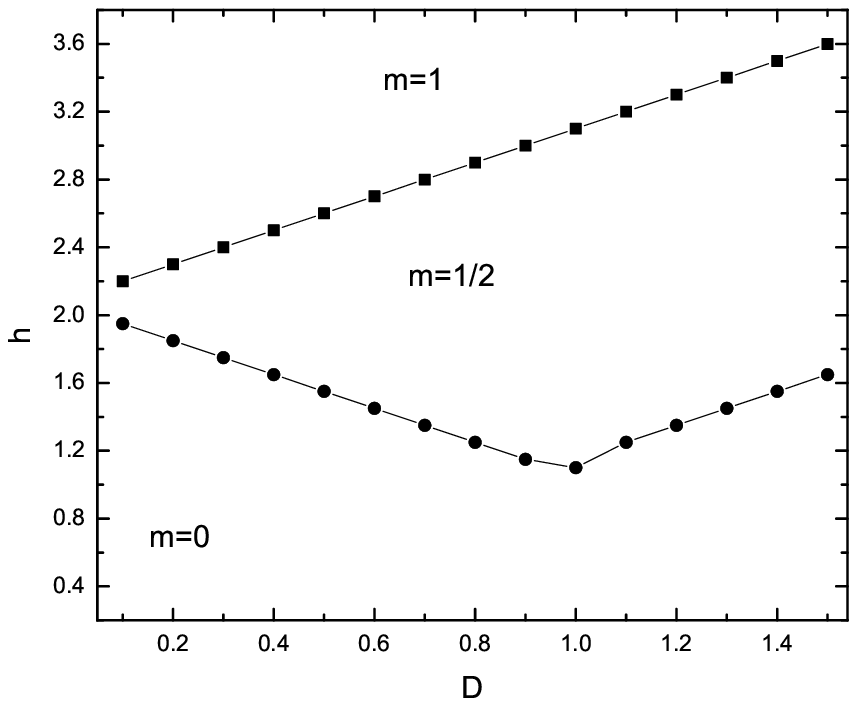}
\caption{\label{fig:Fig2} Magnetization phase diagram of the
ground state of antiferromagnetic Ising chain with single-ion
anisotropy under finite magnetic field. The circle dot-line and
the square-dot line represent the initial field and the saturated
field, respectively.}
\end{center}
\end{figure}
\begin{figure}
\begin{center}
\subfigure[\hspace{0.4cm}]{\label{fig:sub:a}
\includegraphics[width=8.0cm,height=8.0cm,angle=0]{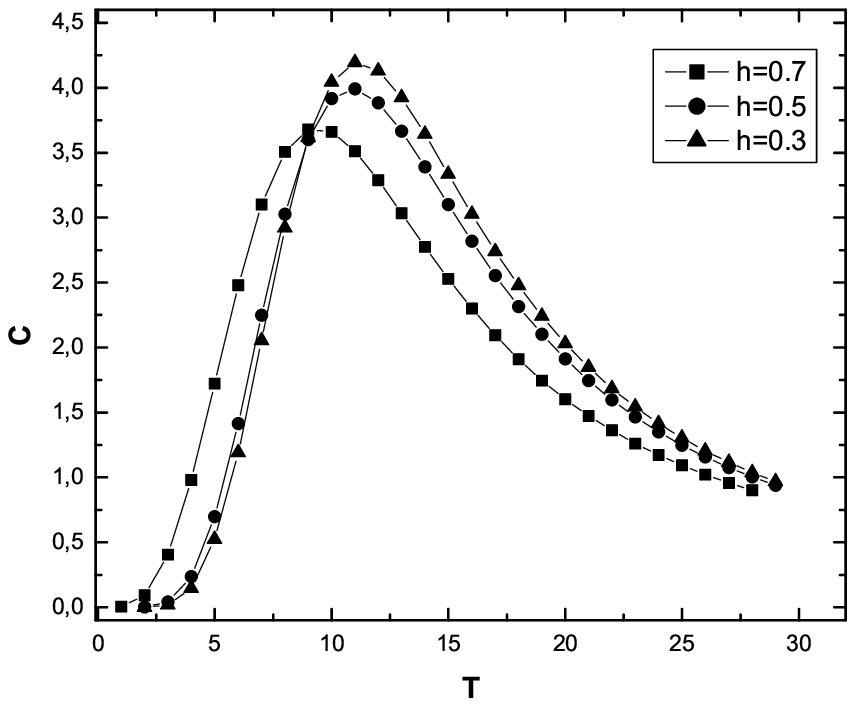}}
\hspace{0.4cm} \subfigure[\hspace{0.4cm}]{\label{fig:sub:b}
\includegraphics[width=8.0cm,height=8.0cm,angle=0]{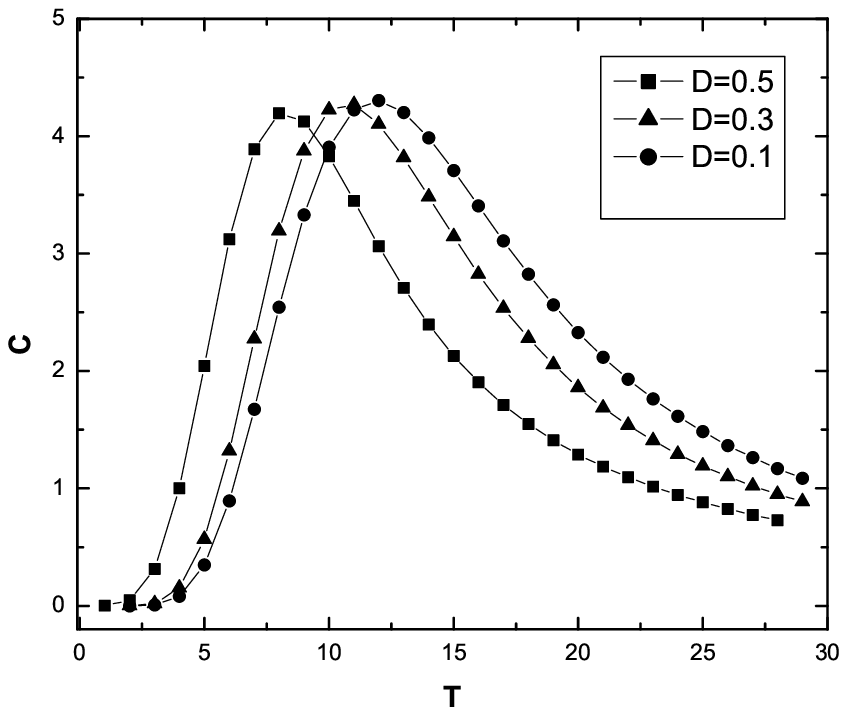}}
\caption{The specific heat $C$ as a function of the temperature
$T$ (unit by $J$): (a) $h=0.3, 0.5, 0.7$ for fixed value of
$D=0.2$ (b) $D=0.1, 0.3, 0.5$ for fixed value of $h=0.1$,
respectively.} \label{fig:sub:Fig1-2}
\end{center}
\end{figure}

It suggested that single-ion anisotropy plays a significant role
for magnetic plateau.\cite{4} In order to examine the effect of
single-ion anisotropy on the magnetization plateaus, the
magnetization $m$ was calculated at finite $h$ for a series of
values of $D$ ($0.1\leq{D}\leq1.5$), and the data were plotted in
Fig. 2. For $D>0.0$, three plateau lines placed at $m=0.0$, $m
=0.5$, and $m=1.0$ are divided by the initial field and saturated
field lines as shown in Fig. 2. The longitudinal coordinate of the
circular-dot line is the beginning point of the field for the
appearance of the plateau $m=0.5$ which corresponds to critical
field $h_{c}$ values, and its ending point is signed by the
square-dot line which corresponds to the saturated field $h_{s}$
values. In addition, the distance between both the lines for the
same value of $D$ is the width of the plateau $m=0.5$. For
$D\geq{1.0}$, the value of saturated field $h_s$ increases with
the increase of $D$. The initial field $h_c$ decreases with the
increase of $D$ in the interval $0.0<D<1.0$  and increases for
around $D>1.0$. Almost same magnetization phase diagram for spin-1
has been carried out using Monte Carlo method with finite spin in
Ref.4.

\begin{figure}
\includegraphics[width=8.0cm,height=8.0cm,angle=0]{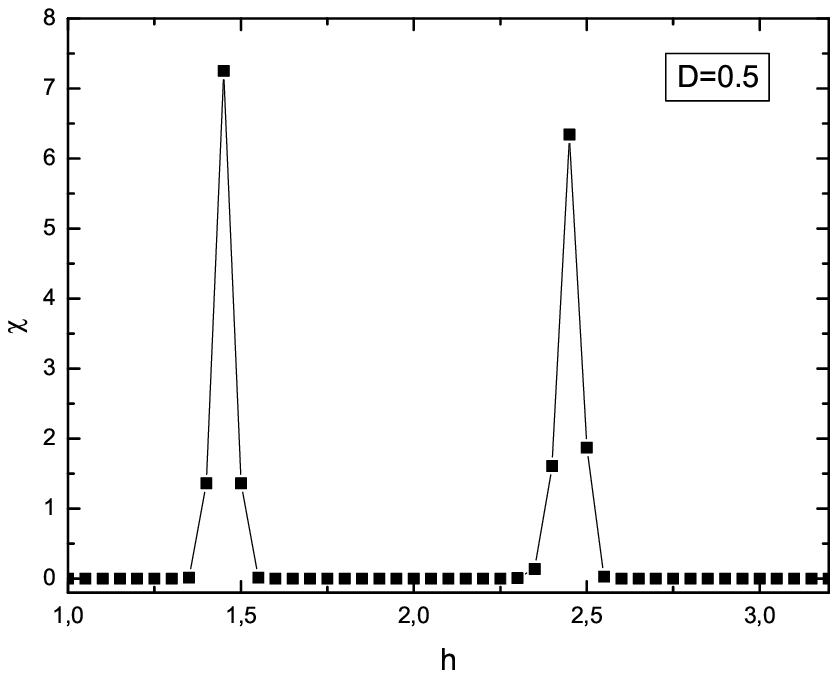}
\caption{\label{fig:Fig4} Susceptibility as a function of magnetic
field $h$ for $D=0.5$ and $T=0.01$ (unit by $J$).}
\end{figure}

Figure 3 (a) and (b) show the specific heat $C$ plotted as a
function of temperature $T$, for fixed value of single-ion
anisotropy field $D=0.2$ at various values of external field
$h=0.3, 0.5, 0.7$, and for fixed value of $h=0.1$ at various
values of $D=0.1, 0.3, 0.5$, respectively. These curves show a
board maximum as Schottky-like round hump about between $T=8$ and
13. But, it is no means that phase transition occurs in
one-dimensional system. Schottky-like round hump in the specific
heat probably reflects energy of the system. For several higher
external fields and single-ion anisotropy fields, at fixed value
of $D$ and $h$, respectively, the peaks get smaller and move
towards zero temperature. It seems clear that the peaks are
related to the magnetization phase diagram of the ground state of
AF Ising chain with single-ion anisotropy under finite magnetic
field. Furthermore, the specific heat curves of the spin chain
which have obtained as a function of temperature are in good
agreement with the experimental data for the compounds
Ni(C$_2$H$_8$N$_2$)Ni(CN)$_4$,\cite{22,23,24,25}
Ni(C$_{10}$H$_8$N$_2$)$_2$Ni(CN)$_4$H$_2$O,\cite{22,23,24} and the
theoretical studies \cite{4,27,28} in literature, respectively.

An other way is to understand the ground state behavior of the
system is investigation field dependence of the susceptibility. In
this reason, the susceptibility were plotted in Fig. 4 as a
function of the magnetic field. It is seen that there are two
peaks in the susceptibility for $D=0.5$ at $T=0.01$. First peak
occur at initial field value, and second peak occur saturated
field value for fixed value of $D$. Both of the peaks indicate the
critical fields where magnetization plateaus appear. Similar two
peaks have been observed in experimental study for compound
[Ni$_2$(Medpt)$_2$($\mu$-ox)(H$_2$O)$_2$](ClO$_4$)$_2$2H$_2$O.\cite{15}

\begin{figure}
\includegraphics[width=8.0cm,height=8.0cm,angle=0]{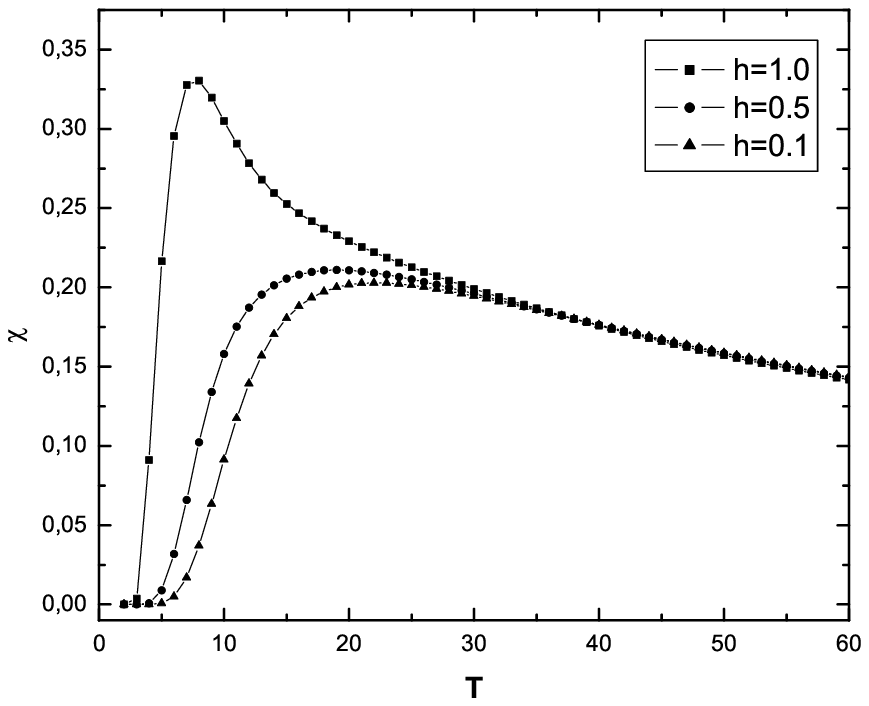}
\caption{\label{fig:Fig5} Susceptibility as a function of
temperature for various $h$ and $D=0.5$ (unit by $J$).}
\end{figure}

On the other hand, for different values of the magnetic field, the
susceptibility was plotted as a function of the temperature $T$
for fixed value of $D=0.5$ in Fig. 5. The susceptibility curve for
$h=1.0$, marked by square-dot line, has relatively a sharp peak
around $T=5$. However, it has a broad maximum for $h=0.5$, and
$h=0.1$ which are denoted by circular-dot and triangular-dot line,
respectively. We roughly say that susceptibility curves decay
exponentially with increasing temperature by the following
relation $\chi\left(T\right)
\approx\exp\left(-E_{g}/k_{B}T\right)$. This characteristic
behavior can be interpreted that AF spin-1 chain with the
single-ion anisotropy has the Haldane gap like Heisenberg AF
systems. Our results clearly compatible with the experimental
results for compounds
[Ni$_2$(Medpt)$_2$($\mu$-ox)(H$_2$O)$_2$](ClO$_4$)$_2$2H$_2$O,\cite{15}
[Ni$_{2}$(Medpt)$_{2}$($\mu$-ox)($\mu$-N$_{3}$)](ClO$_{4}$)0.5H$_{2}$O,
\cite{16} Ni(C$_5$H$_{14}$N$_2$)$_2$N$_3$(PF$_6$),\cite{27} and
3,3',5,5'-tetrakis (N-tert-butylaminxyl) biphenyl,\cite{17} and
theoretical results \cite{27,28} in literature. However, in the
present study, the specific heat and magnetic susceptibility
curves which obtained as a function of temperature have not a
spin-Peierls transition. Because in the compounds synthesized
experimentally, a spin-Peierls transition in $C$ and $\chi$ at low
temperatures are due to a small amount of magnetic impurities and
derivations from stoichiometry.\cite{34}

In the majority of plateau mechanism which have been proposed up
to now the purely quantum phenomena play a curial role. The
concepts of magnetic quasi-particles and the strong quantum
fluctuations are regarded to be first important for understanding
of these processes. Particularly, for a number of systems it was
shown that the plateaus are caused by the presence of the spin gap
in the spectrum of magnetic excitations in the external field.
However, we have used quasi-classical Hamiltonian (1) in this
study, and magnetic plateaus, magnetic phase diagram, specific
heat, magnetic susceptibility of 1D AF spin-1 chain have been
obtained. Obtaining results clearly consistent with experimental
and theoretical results which ensure that quasi-classical approach
can be used to examine the ground state properties of 1D AF spin-S
gapped systems as spin-1.

Finally, we conclude that the single-ion anisotropy in the Ising
model has a significant effect on the magnetic properties under
the external field. Therefore, we studied this phenomenon in the
one-dimensional spin-1 AF Ising chain with single-ion anisotropy.
When the external field varied from the zero to the saturated
field, the 2S+1 step-like plateaus occurred owing to the
co-existence of AF interaction of ($S_{i}^z S_{i+1}^z$) and
positive single-ion anisotropy $(S_{i}^z)^2$. Hence, the energy
levels, separated by the crystal field (i.e. single-ion
anisotropy) in the system of two particles, correspond to the
plateaus in the ground state.

We acknowledge useful discussion with Hamza Polat.

\end{document}